\documentclass[
    reprint,
    aps,
    prx,
    superscriptaddress,
    longbibliography,
    floatfix
]{revtex4-2}

\usepackage{graphicx}   
\usepackage{dcolumn}    
\usepackage{bm}         
\usepackage{amsmath}
\usepackage{amssymb}
\usepackage{amsfonts}

\usepackage{physics}
\usepackage{siunitx}

\usepackage{layouts}
\usepackage{svg}

\usepackage[normalem]{ulem}
\usepackage{soul}
\usepackage{xcolor}

\usepackage[utf8]{inputenc}

\usepackage[
    colorlinks=true,
    linkcolor=blue,
    citecolor=blue,
    urlcolor=blue
]{hyperref}

\AtBeginDocument{\RenewCommandCopy\qty\SI}
\begin{document}

\preprint{APS/123-QED}

\title{Mixed spin-boson coupling for qubit readout with suppressed residual shot-noise dephasing}

\author{Jinlun Hu}
\thanks{These authors contributed equally.}
\email{J.Hu-4@tudelft.nl}
\affiliation{
    QuTech and Kavli Institute of Nanoscience, Delft University of Technology, Delft 2600 GA, The Netherlands
 }

\author{Antonio L.~R.~Manesco}%
\thanks{These authors contributed equally.}
\email{am@antoniomanesco.org}
\affiliation{%
    Kavli Institute of Nanoscience, Delft University of Technology, Delft 2600 GA, The Netherlands
}%

\affiliation{
    School of Applied and Engineering Physics, Cornell University, Ithaca, NY, 14853, USA
}

\author{André Melo}

\affiliation{%
    Kavli Institute of Nanoscience, Delft University of Technology, Delft 2600 GA, The Netherlands
}%

\author{Taryn V. Stefanski}
\thanks{Present address: QphoX, 2628 XG, Delft, The Netherlands}
\affiliation{
    QuTech and Kavli Institute of Nanoscience, Delft University of Technology, Delft 2600 GA, The Netherlands
 }
 \affiliation{Quantum Engineering Centre for Doctoral Training, H. H. Wills Physics Laboratory and Department of Electrical and Electronic Engineering, University of Bristol, BS8 1FD, Bristol, UK}

\author{Christian Kraglund Andersen}
\email{C.K.Andersen@tudelft.nl}
\affiliation{
    QuTech and Kavli Institute of Nanoscience, Delft University of Technology, Delft 2600 GA, The Netherlands
}

\author{Valla Fatemi}
\email{vf82@cornell.edu}
\affiliation{
    School of Applied and Engineering Physics, Cornell University, Ithaca, NY, 14853, USA
}

\date{July 24, 2026}

\begin{abstract}
Direct dipole coupling between a two-level system and a bosonic mode describes the interactions present in a wide range of physical platforms. 
In this work, we study a coupling that is mixed between \textit{two} pairs of quadratures of a bosonic mode and a spin. 
In this setting, we can suppress the dispersive shift while retaining a nonzero Kerr shift, which results in a quadratic relationship between shot-noise dephasing and thermal photons in the oscillator.
We demonstrate this configuration with a simple toy model, quantify the expected improvements to photon shot-noise dephasing of the spin, and describe an approach to fast qubit readout via the Kerr shift. 
Further, we show how such a regime is achievable in superconducting circuits because magnetic and electric couplings can be of comparable strength, using two examples: the Cooper pair transistor and the fluxonium molecule.

\end{abstract}

\maketitle

\section{\label{sec:intro}Introduction}

Within the field of quantum technologies, it is imperative that we build qubit systems that allow convenient methods to manipulate and readout the qubits~\cite{nielsen2010quantum, preskill2018quantum, kjaergaard2020superconducting, krantz2019quantum}. In many cases, these controls are achieved by coupling the qubit system coherently with bosonic modes~\cite{haroche2006exploring, blais2004cavity, blais2021circuit}. For readout in particular, qubits are often coupled to a readout mode via a dipole-like coupling $\sigma_i (a^{\dagger} \pm a)$, where $i\in x,y$. 
The dipole coupling can manifest as a coupling of the quadratures $X=a^{\dagger} +a$ or $ P= i(a^{\dagger} - a)$  to either the qubit operator $\sigma _x$ or to the operator $\sigma_y$, depending on its physical implementation. 
For example, superconducting qubits may be capacitively or inductively coupled to a readout resonator~\cite{chen2023transmon, dassonneville2020fast, heinsoo2018rapid, chen2012multiplexed, hinderling2024flip}. 
In these platforms, readout is commonly performed in the dispersive regime, where the resonator frequency shifts depending on the qubit state~\cite{wallraff2005approaching, walter2017rapid}.
Within perturbation theory, this response manifests in the Hamiltonian as $\chi \sigma_z a^\dagger a$.
Thereby, the nature of the coupling (i.e., as arising from coupling either $X$ or $P$ of the resonator to either $\sigma_x$ or $\sigma_y$ of the qubit, or combinations thereof) is hidden. 

The dispersive frequency shift also limits qubit performance. 
The qubit reciprocally picks up a frequency shift proportional to the number of photons in the resonator, also known as the ac-Stark shift~\cite{schuster2005ac}.
This means that the residual average photon number $\bar{n}_{\mathrm{th}}$ in the resonator will induce random shifts in the qubit frequency, leading to qubit dephasing rates that are linearly proportional to $\bar{n}_{\mathrm{th}}$~\cite{gambetta2006qubit, clerk2007using, rigetti2012superconducting, 2019_Wang_PRAppl_11}.
This dephasing channel is referred to as photon shot noise. 
For superconducting qubits, including transmon qubits and fluxonium qubits, which have intrinsic protection against charge dispersion and are linearly protected to flux noise at sweet spots, photon shot noise is a limiting factor in coherence times~\cite{2019_Wang_PRAppl_11,somoroff2023millisecond}.
For example, to demonstrate large coherence times for fluxonium qubits, the dispersive shift $\chi$ is often designed to be much smaller than the readout resonator linewidth in order to protect the qubit from photon shot noise~\cite{2019_Nguyen_PRX_9}.
This has the effect of requiring longer readout times to achieve desired measurement fidelities. 
Previous works with superconducting qubits have also studied how adding extra nonlinear elements can circumvent some of the downsides of conventional dispersive readout, such as reducing measurement-induced state-transition or Purcell decay~\cite{chapple2025balanced, wang202499, sunada2024photon, dassonneville2020fast, govia2017enhanced}. However, these methods still rely on an effective dispersive Hamiltonian term that is intrinsically sensitive to photon shot noise. A complementary strategy is to suppress photon-shot-noise dephasing by reducing the dispersive shift itself. Zhang \textit{et al.}~\cite{zhang2017suppression} demonstrated an approach using a tunable-coupling qubit, in which two hybridized qubit modes contribute oppositely to the cavity pull, enabling cancellation of the dispersive shift and suppression of photon-shot-noise dephasing. However, because the same cancellation eliminates conventional dispersive readout contrast, the readout required mapping the qubit state to an auxiliary excited state.

\begin{figure}[t]
    \centering
    \includegraphics[width=\linewidth]{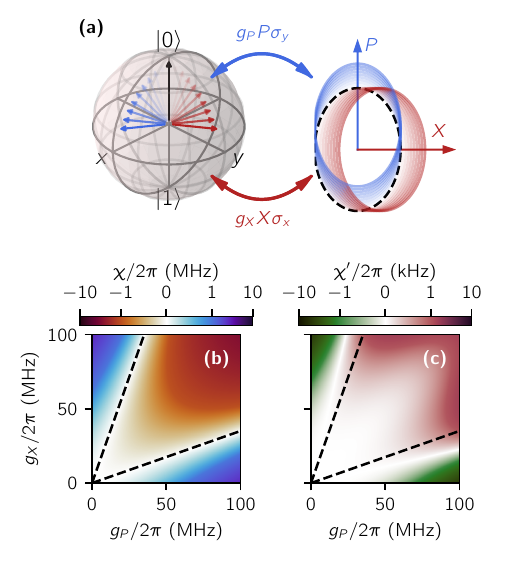}
    \caption{
        (a) Depiction of the mixed-coupling model.
        The Bloch sphere on the left shows the time evolution of the $\ket{0}$ state in the presence of the two couplings $g_X$ (red) and $g_P$ (blue).
        On the right, we show the corresponding phase space displacement of the harmonic oscillator for each coupling term.
        Symlog colormaps of dispersive (b) and Kerr (c) shifts as a function of the couplings $g_X$ and $g_P$.
        Here, we use $\omega_q/(2\pi) = \SI{5}{\giga\hertz}$ and $\omega_r/(2\pi) = \SI{8}{\giga\hertz}$ so that $|\omega_q - \omega_r|\gg g_X, g_P$.
        The dashed lines indicate the analytical $\chi=0$ condition from Eq.~(\ref{eq:analytical_chi}).
        We show the $\chi=0$ condition also in panel (c) to indicate that these two corrections are not simultaneously suppressed.
    }
    \label{fig:tls}
\end{figure}

In this work, we explore the scenario in which a qubit is coupled to a resonator mode through a mixed coupling, which in superconducting circuits can be achieved using a combination of inductive and capacitive elements~\cite{blais2007quantum,lu2017universal, wang202499}.
Specifically, we pair $\sigma_x$ and $\sigma_y$ with distinct quadratures of a bosonic mode, see also Fig.~\ref{fig:tls}, and demonstrate that a mixed coupling can provide a suppression of the dispersive shift. We find that the photon shot noise rate scales as the square of the residual photon number ($\sim\bar{n}_{\mathrm{th}}^2$) in this regime, implying significant improvements relative to usual dispersive coupling for typical $\bar{n}_{\mathrm{th}}\sim 0.01$.
Inspired by this result, we propose a nonlinear readout scheme by measuring the next available term in perturbation theory, the Kerr shift $\chi^{\prime} \sigma_z a^{\dagger} a^{\dagger} aa$, and show that this readout is competitive in both speed and fidelity.
We show how to achieve mixed coupling in superconducting qubit circuits, making them ideal platforms for non-linear readout.
By presenting a method that suppresses residual shot noise without the need for extra parametric drives, our work adds new features to readout that may support superconducting quantum circuits with increased performance.

\section{\label{sec:theory}Mixed coupling and nonlinear readout}

We first consider a minimal model, which consists of a qubit coupled to a resonator with the Hamiltonian
\begin{equation}
    H = \hbar\frac{\omega_q}{2}\sigma_z + \hbar\omega_r a^{\dagger} a + H_{\text{coupling}}~,
    \label{eq:generic_spin_boson}
\end{equation}
where $\sigma_i$ are Pauli matrices acting in the qubit degrees of freedom, and $a^{\dagger}$ are creation and $a$ are annihilation operators of the bosonic field, $\omega_q$ is the energy separation between the qubit levels, and $\omega_r$ is the frequency of the resonator.
We consider a mixed coupling between the qubit and the resonator with the Hamiltonian
\begin{equation}
    H_{\text{coupling}} = \hbar g_X \sigma_x X + \hbar g_P \sigma_y P~,
    \label{eq:coupling}
\end{equation}
with $X = a^{\dagger} + a$ and $P = i(a^{\dagger} - a)$.
We remark that a realization of such a coupling where $g_X$ and $g_P$ are independently tunable allows obtaining a lab-frame Jaynes-Cummings Hamiltonian when $g_X = - g_P$, while an anti-Jaynes-Cummings Hamiltonian is achieved with $g_X = g_P$.

In the weak-coupling regime, $|g_X|, |g_P| \ll |\omega_r - \omega_q|$, we perform a Schrieffer-Wolff transformation that keeps non-rotating wave approximation (non-RWA) effects (see Appendix \ref{app:lowdin}).
We then find that, within leading order in perturbation theory, the dispersive shift is
\begin{equation}
    \chi = \frac{1}{\omega_r^2 - \omega_q^2}\left[4 \omega_r g_P g_X -2\omega_q (g_X^2 + g_P^2) \right ]~.
    \label{eq:analytical_chi}
\end{equation}
This expression is not monotonic in $g_X$ and $g_P$; thus, we see that it is possible to completely suppress the dispersive shift for non-zero $g_X$ and $g_P$, as shown in Fig.~\ref{fig:tls} (b).
Moreover, we also see from the Schrieffer-Wolff transformation that we get non-linear corrections to the Hamiltonian in the form of
\begin{align}
    H_{Kerr} =\hbar \chi'\sigma_z\, a^\dagger a^\dagger a a, 
\end{align}
where we denote $\chi'$ as the Kerr shift, i.e., a qubit-state-dependent Kerr coefficient, see Fig.~\ref{fig:tls} (c).

The nonlinear interaction term $\chi' \sigma_z a^\dagger a^\dagger a a$ encodes the information about the qubit state in the change of resonator anharmonicity. More specifically, the Kerr shift $\chi'$ can be written as
\begin{equation}
    \chi^\prime =(K_{r,\ket{1}} - K_{r, \ket{0}})/4, 
\end{equation}
where $K_{r,\ket{q}}$ denotes the resonator anharmonicity when the qubit is $\ket{q}$ state.
In this section, we demonstrate how we can use the Kerr shift to obtain efficient readout without suffering from residual shot noise.

\begin{figure}
    \centering
    \includegraphics[width=\linewidth]{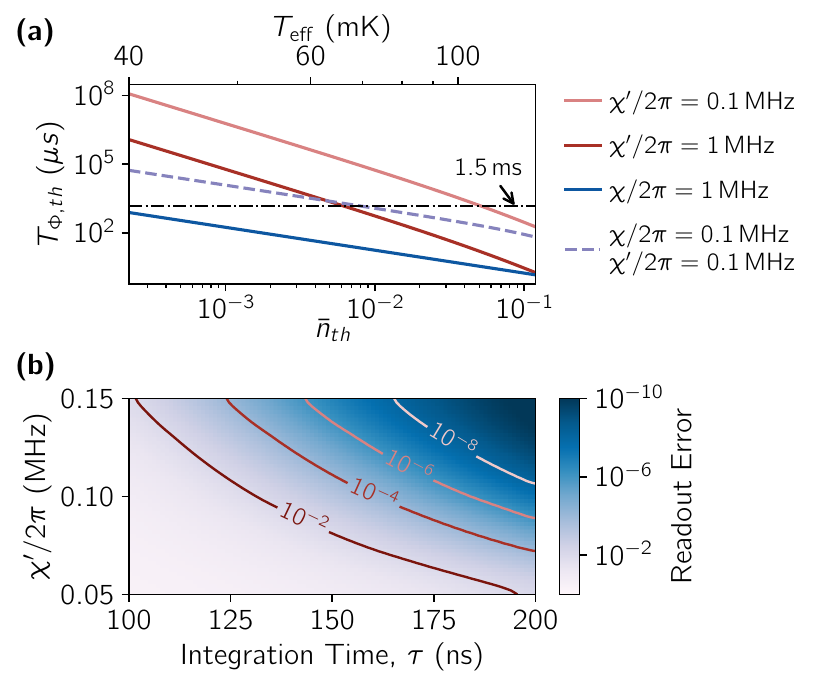}
    \caption{
    Nonlinear readout.
    (a) Qubit dephasing time limit due to thermal photon shot noise based on linear (blue) and nonlinear (light and dark red) responses, based on Eqs.~\eqref{eq:linear_dephasing} and~\eqref{eq:Gamma_final_4th}. 
    The purple dashed curve shows a result when both $\chi$ and $\chi'$ are present. We use $\kappa/2\pi = 3$ MHz and $\omega_r/2\pi = 7$ GHz.
    The black dot-dashed line indicates the best qubit coherence measured for superconducting qubits with $T_2^* = 1.5$ ms~\cite{somoroff2023millisecond}.
    (b) Readout error of the nonlinear readout scheme as a function of integration time for different values of $\chi'$.
    Drive amplitude is fixed such that the peak transient photon number in the resonator is less than $15$.
    The quantum efficiency is assumed to be 100\%, and we use $\kappa/2\pi = 3$ MHz. 
    }
    \label{fig:nonlinear_readout}
\end{figure}

\subsection{Photon shot noise}
First, we consider photon shot noise.
For the conventional dispersive interaction with the linear term $\chi \sigma_z a^\dagger a$, the shot noise dephasing is proportional to $\bar{n}_{\mathrm{th}}$ in the limit of $\bar{n}_{\mathrm{th}}\ll1$~\cite{2019_Wang_PRAppl_11, clerk2007using}:
\begin{equation}
\Gamma^{th}_{\phi,\text{linear}} = \frac{4\kappa \chi^2}{\kappa^2 + 4\chi^2}\bar{n}_{\mathrm{th}},
    \label{eq:linear_dephasing}
\end{equation} 
where $\kappa$ is the resonator linewidth. In thermal equilibrium, $\bar{n}_{\mathrm{th}}$ is determined by the resonator frequency $\omega_r$ and the effective resonator temperature $T_{\text{eff}}$ through $\bar{n}_{\mathrm{th}} = 1/(e^{\hbar \omega_r/k_B T_\text{eff}}-1)$. In some recent experimental setups, $\bar{n}_{\mathrm{th}}$ has been measured at different frequencies to be in the order of $10^{-3}$, indicating that $T_{\text{eff}}$ is around $50$ mK~\cite{somoroff2023millisecond, 2019_Wang_PRAppl_11}. To avoid being limited by photon shot noise and achieve record-high qubit coherence, the $\chi/\kappa$ ratio is sometimes chosen to be very small, compromising the readout speed~\cite{somoroff2023millisecond}.

In contrast to linear dispersive coupling, the nonlinear interaction leads to a dephasing rate that is quadratic in
$\bar n_{\mathrm{th}}$ in the low-photon limit. Specifically, we derive the dephasing rate from a master equation in the presence of thermal photon populations, see Appendix~\ref{app:dephasing}. Assuming $\kappa \gg\chi'$, we find the dephasing rate to be

\begin{equation}
\Gamma^{th}_{\phi,\text{nl}}
=
\frac{8\chi'^2}{\kappa}\,
\bar{n}_{\mathrm{th}}^2(\bar{n}_{\mathrm{th}}+1)(9\bar{n}_{\mathrm{th}}+1).
\label{eq:Gamma_final_4th}
\end{equation}

In the low-photon limit,
$\bar{n}_{\mathrm{th}}\ll1$, this simplifies to a quadratic dependence on $\bar{n}_{\mathrm{th}}$

\begin{equation}
    \Gamma^{th}_{\phi,\text{nl}} = \frac{8 {\chi'}^2}{\kappa} {\bar{n}_{\mathrm{th}}^2},
    \label{eq.nonlinear_dephasing}
\end{equation} and therefore largely suppressed since $\bar{n}_{\mathrm{th}} \ll 1$.

Fig.~\ref{fig:nonlinear_readout}(a) compares qubit dephasing times attributed to the linear and nonlinear couplings according to Eq.~\eqref{eq:linear_dephasing} and~\eqref{eq:Gamma_final_4th}. 
The qubit dephasing time limit $T_{\phi,th} = 1/\Gamma^{th}_{\phi}$ is lifted to above $20$ ms at around $50$ mK if $\chi=0$ and $\chi'< 1$ MHz.
Because of the dramatic differences in scaling with $\bar{n}_{\mathrm{th}}$, we expect that residual $\chi \neq 0 $ due to design or fabrication imperfections would ultimately limit improvements to dephasing.
Therefore, the main engineering lever arm becomes the quadratic scaling of dephasing with $\chi$ for $\chi \ll \kappa$.

\subsection{Readout}
\label{sec:readout}
To simulate the resonator response as we send in a readout signal at the resonator frequency, we consider the semi-classical equation of motion~\cite{2020_Andersen_PRApplied_13, stefanski2024flux, stefanski2024improved}
\begin{equation}
    \dot{\alpha} = - i \sigma_z (\chi + 2\chi' |\alpha|^2) \alpha - \frac{1}{2} \kappa \alpha - \sqrt{\kappa}\alpha_\mathrm{in},
    \label{langevin_resonator}
\end{equation}
where $\alpha$ is a complex number that describes the amplitude of a coherent state in the resonator, $\alpha_\mathrm{in}$ is the amplitude of the input field $\alpha_\mathrm{in} = - \epsilon / \sqrt{\kappa}$, and $\epsilon$ is the driving amplitude of the input field. In this equation, we include both the linear and non-linear responses. Note that we simulate this equation semi-classically, meaning that $\sigma_z$ takes a discrete value of $-1$ or $+1$ for the qubit states $\ket{0}$ and $\ket{1}$, respectively. The output field satisfies $\alpha_\mathrm{out} = \alpha_\mathrm{in} + \sqrt{\kappa}\alpha$, and the difference in the output field for the qubit in $\ket{0}$ and $\ket{1}$, together with the corresponding optimal matched-filter weight, readily gives us the readout signal-to-noise ratio~\cite{gardiner1985input, stefanski2024flux, sank2025system, didier2015fast, didier2015heisenberg}. 

We also note that the semi-classical approach that we employ here explicitly assumes that the resonator is occupied by a coherent state. However, for large nonlinearities, the resonator field may become distorted~\cite{kirchmair2013observation}. On the other hand, the semiclassical approach works well if the nonlinearity is significantly smaller than the resonator linewidth~\cite{2020_Andersen_PRApplied_13}.
Thus, we explicitly design our protocol to operate in the regime of small self-Kerr. In Sec.~\ref {sec:fxm}, we will verify this semi-classical approach further with a comparison to a full master equation simulation.

To study the resonator response from the contribution of only the nonlinear term, we take $\chi = 0$. For each value of $\chi'$, we choose the drive amplitude $\epsilon$ such that the maximum transient resonator photon population within the integration time is below 15. We note that within this model, for a single value of $\epsilon$, the resonator photon number $n$ does not differ for the qubit in $\ket{0}$ or $\ket{1}$. 
Crucially, once $|\alpha|$ exceeds 1, the resonator field picks up a quadratically increasing separation in its imaginary part, thanks to the $\chi'$ term. The results, as shown in Fig.~\ref{fig:nonlinear_readout}(b), confirm that with a moderate Kerr shift amplitude of $0.1$ MHz, the readout error can go below $10^{-4}$ within 170 ns.
We remark that these steady-state calculations have not taken advantage of any readout pulse optimization methods that have been useful for shortening readout times in usual linear dispersive readout settings~\cite{mcclure2016rapid, walter2017rapid, boutin2017resonator}.

In summary, we demonstrated a readout scheme that takes advantage of the quadratic relation in the term $\chi' \sigma_z a^\dagger a^\dagger a a$. Additionally, we showed that this coupling scheme manifests a suppressed qubit dephasing rate when the residual photon number in the resonator is small. Next, we will show how to realize mixed coupling with real circuits such that we are able to suppress $\chi$ without suppressing $\chi'$.

\section{\label{sec:ciruit_realiz}Circuit realizations of mixed-coupling}

\subsection{\label{sec:cpt}Cooper pair transistor}
\begin{figure}[!t]
    \centering
    \includegraphics[width=\linewidth]{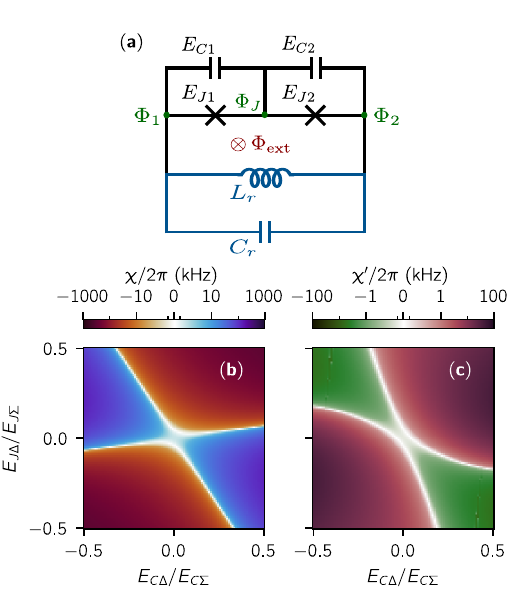}
    \caption{
    (a) Circuit diagram of the Cooper pair transistor. Here, $\Phi_1$, $\Phi_J$, and $\Phi_2$ denote physical node-flux variables, while $\Phi_{\mathrm{ext}}$ denotes the externally applied loop flux. 
    Symlog colormaps of numerically calculated dispersive shift (b) and Kerr shift (c) with $E_{C_r}=2\pi\times\SI{10}{\giga\hertz}$, $E_{L_r}=2\pi \times\SI{100}{\giga\hertz}$, $E_{J\Sigma} = E_{J1} + E_{J2} =2\pi\times\SI{18}{\giga\hertz}$, $E_{C\Sigma} = E_{C1} + E_{C 2}= 2\pi\times\SI{10}{\giga\hertz}$.
    With these parameters the resonator frequency is $\omega_r \approx 2\pi\times\SI{63}{\giga\hertz}$ while the qubit frequency is $\omega_q \sim 2\pi\times \SI{5}{\giga\hertz}$. 
    }
    \label{fig:CPT_circuit_and_results}
\end{figure}

Our first circuit implementation of the mixed-coupling Hamiltonian is shown in Fig.~\ref{fig:CPT_circuit_and_results}(a). In the Cooper-pair box limit, the qubit-resonator coupling in this circuit contains the two terms in Eq.~(\ref{eq:coupling}), see also Appendix~\ref{app:cpt-derivation} for the full derivation and the complete expression. Thus, we obtain the Hamiltonian
\begin{align}
    \mathcal{H}_{\text{coupling}} \sim -
    \frac{2E_{C\times}|z|}{\varepsilon}n_R\sigma_x
        +
    \frac{E_{J\Delta}}{4|z|}\phi_R\sigma_y.
    \label{eq:cpt_coupling}
\end{align}
where $\phi_R$ is the resonator mode phase coordinate, $n_R$ is its conjugate variable, and 
\begin{align}
    z \equiv -\cos\frac{\phi_\text{ext}}{2} + i \frac{E_{J\Delta}}{E_{J\Sigma}} \sin\frac{\phi_\text{ext}}{2}~,
    \\
    m\equiv
    \frac{4E_{CI}n_g'}{E_{J\Sigma}}, 
    \quad
    \varepsilon
    \equiv
    \sqrt{m^2+|z|^2}.
    \label{eq:cpt_convenient_definitions}
\end{align}
$n_g'=2n_g-1$ is the dimensionless detuning from the
charge-degeneracy point, and $\phi_\mathrm{ext}\equiv\Phi_{\mathrm{ext}}/\varphi_0$ is the dimensionless external phase.

The energy parameters are related to the circuit parameters shown in Fig.~\ref{fig:CPT_circuit_and_results}(a) as
\begin{align}
    E_{J\Sigma}=E_{J1}+E_{J2},
    \qquad
    E_{J\Delta}=E_{J1}-E_{J2}.
    \\
    E_{C\Sigma}=E_{C1}+E_{C2},
    \qquad
    E_{C\Delta}=E_{C1}-E_{C2}.
\end{align}
The effective charging energies $E_{CI}$ and $E_{C\times}$ are
obtained from the inverse capacitance matrix, as defined in
Appendix~\ref{app:cpt-derivation}. In particular,
\begin{equation}
    E_{C\times}
    =
    -\frac{e^4}{4\det(\mathbf C)}
    \frac{E_{C\Delta}}
    {E_{C1}E_{C2}}.
\end{equation}

From Eq.~(\ref{eq:cpt_coupling}), we conclude that up to a rotation of the transverse qubit axes, the
Josephson-energy asymmetry $E_{J\Delta}$ controls the
$X$-quadrature coupling $g_X$, while the capacitive asymmetry,
through the effective cross-charging energy $E_{C\times}$,
controls the $P$-quadrature coupling $g_P$.

To verify the influence of the additional coupling terms in Eq.~(\ref{eq:cpt-interaction-energy-basis}) and higher qubit states, we compute the dispersive and Kerr shifts numerically using the Python package \texttt{scQubits}~\cite{groszkowski2021scqubits}. Since some of the coupling terms in Eq.~(\ref{eq:cpt-interaction-energy-basis}) are as large as the qubit frequency ($\omega_q \sim E_{J\Sigma}$), we fix a large resonator frequency $\omega_r \gg \omega_q$ to stay in the perturbative regime.
This choice is made to illustrate a mixed coupling in a circuit realization, and we devote Sec.~\ref{sec:fxm} for realistic circuit simulations.
In Fig.~\ref{fig:CPT_circuit_and_results} (b), we demonstrate the suppression of the dispersive shift by tuning $E_{C\Delta}$ and $E_{J\Delta}$ despite the additional complexity of the circuit Hamiltonian.
Moreover, the nonlinear readout remains possible since the Kerr shift remains nonzero, as shown in Fig.~\ref{fig:CPT_circuit_and_results}(c).

\begin{figure*}[t]
    \centering
    \includegraphics[width=\linewidth]{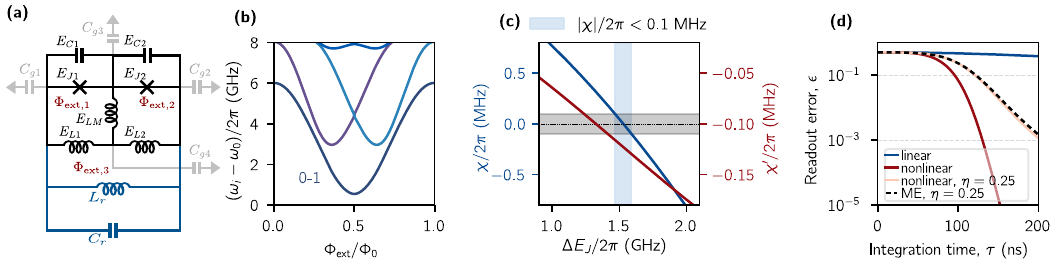}
    \caption{
        (a) Circuit diagram of a fluxonium molecule (in black) galvanically connected to a readout resonator (in blue). For all circuit simulations here, $\Phi_{\text{ext,3}}=0$.
        (b) Numerically simulated energy spectrum of an asymmetric fluxonium molecule with the following energy parameters: $E_{J1} /2\pi = 3.63$ GHz, $E_{J2} /2\pi = 5.17$ GHz, $E_{L1} = E_{L2} = E_{LM} = 2\pi \times 1.2$ GHz, $E_{C1}/2\pi = 3.08$ GHz, $E_{C2} /2\pi = 2.52$~GHz. $\Phi_\text{ext} = (\Phi_{\text{ext,1}}+\Phi_{\text{ext,2}})/2$.
        The four nodes of the fluxonium are capacitively coupled to ground with $C_{g1}=C_{g2}=C_{g3}=C_{g4}=0.1$~fF. (c) Shifts $\chi$ and $\chi'$ from circuit numerical simulations at $\Phi_{\mathrm{ext}}/\Phi_0=0.5$ as a function of the energy difference between the two Josephson junctions $\Delta E_J = E_{J1}-E_{J2}$. The circuit parameters are the same as in (b) with $E_{J1}/2\pi =E_{J2} /2\pi = 4.4$~GHz when $\Delta E_J=0$.
        The area shaded in blue indicates the regime where $|\chi|/2\pi<0.1$ MHz, which gives rise to at worst $T_{\phi,\text{linear}} = 1.1$~ms using $\kappa/2\pi = 3$ MHz and $T_{\text{eff}} = 65$ mK. 
        (d) Readout error as a function of integration time considering different contributions. The red solid line uses $\chi'/2\pi = 0.12$ MHz. The pink line is plotted with the same $\chi'$ but with quantum efficiency $\eta$ = 0.25. The black dashed line is the result of a full master equation simulation with the same $\chi'$ and $\eta$ = 0.25. For comparison, the blue line shows readout error with only $\chi/2\pi =0.12$ MHz and $100\%$ quantum efficiency. The peak transient photon number is 15 and $\kappa/2\pi = 3$ MHz.
    }
    \label{fig:4 fluxonium molecule}
\end{figure*}

\subsection{\label{sec:fxm}Fluxonium molecule}

The Cooper pair transistor, in the described charging limit, is convenient to understand as it is readily modeled as a two-level system with well-understood eigenstates. 
However, the sensitivity to charge noise means that it is unlikely to serve as a technologically relevant platform.
To avoid detrimental charge noise while retaining a low-lying pair of computational qubit states, we inductively shunt the island to form a fluxonium molecule~\cite{2017_Kou_PRX_7}.
Specifically, we add the three inductive elements to the circuit as shown in Fig.~\ref{fig:4 fluxonium molecule}(a). The fluxonium molecule operates in a regime of $E_{J} > E_{L}$. Thus, the inductive elements should be in the so-called superinductor regime~\cite{masluk2012microwave}. What's more, the charging energy is also larger compared to that of typical single fluxoniums.
The fluxonium molecule can be understood as two individual fluxonium qubits that are strongly coupled near $\Phi_\mathrm{ext} = (\Phi_{\text{ext,1}} + \Phi_{\text{ext,2}})/2 = 0.5 \Phi_0$. At this flux bias, the two lowest-energy states are well separated in energy from the rest of the energy spectrum, see Fig.~\ref{fig:4 fluxonium molecule}(b).
The resulting ground and first excited states are even and odd superpositions of current flowing in opposite directions in their corresponding loops.
The energy spectrum shows high-level transitions in the bandwidth of typical readout frequencies of $5$ to $8$ GHz. For fluxonium qubits, it is well-studied how these higher excited states contribute to the dispersive shift~\cite{zhu2013circuit, stefanski2024flux}.
Similarly, for the fluxonium molecule, the higher excited states will impact both the dispersive shift and the Kerr shift; thus, the full spectrum must be considered in the calculation of the circuit properties.

Here, we compute the dispersive and Kerr shifts numerically with realistic circuit parameters; see the Zenodo repository~\cite{zenodo_code} for the code. Similar to the Cooper pair transistor, the imbalance in capacitive and inductive coupling provides the means to suppress the dispersive shift. As a concrete example, we show in Fig.~\ref{fig:4 fluxonium molecule}(c) that as we increase the imbalance in the two Josephson energies, we reach a point where $\chi$ goes to zero while we obtain $\chi'/2\pi \approx -0.1$~MHz. For these calculations, we have $\omega_q = 107$ MHz and $\omega_r = 6.15$~GHz with a resonator impedance of $100~\Omega$, and resonator anharmonicity $K_r/2\pi = 0.58$~MHz. We note that the somewhat high resonator impedance helps in achieving a larger $\chi'$ without increasing the self-Kerr of the resonator too much; see also Appendix~\ref{app:circ-simulation-details} for more details. In Fig.~\ref{fig:4 fluxonium molecule}(c), we present the actual energy difference between the two Josephson junctions in units of GHz to indicate a window of $|\chi/2\pi|<0.1$~MHz. This window corresponds to around $0.1$ GHz of $\Delta E_J$, while $\chi'$ remains non-zero at around $-0.12$ MHz. In other words, we allow for a relative precision of 2.5\% in the junction energies. Although such precision is challenging on a wafer scale, the junctions for a specific fluxonium molecule are likely to be located within small windows where spread in junction energies below 2\% has been demonstrated~\cite{kreikebaum2020improving}.

Taking the values of $\chi$ and $\chi'$ given by the circuit simulations as shown in Fig.~\ref{fig:4 fluxonium molecule}(c), we perform time-domain readout simulations as outlined in Sec.~\ref {sec:readout}. Under the condition that the maximum transient resonator photon population is below 15, we obtain readout fidelities as shown in Fig.~\ref{fig:4 fluxonium molecule}(d), where the readout error goes below $10^{-4}$ within 150~ns assuming perfect measurement efficiency. For a more realistic value of the measurement efficiency of 25\%, we still obtain a measurement error around $10^{-3}$ within $200$~ns. Using the same conditions, we also plot the result from the full master equation simulation, and we see that the result is very similar to the semiclassical approach. Note that in Fig.~\ref{fig:4 fluxonium molecule}(d), we also plot the readout fidelity contribution from a small $\chi$ set to the same value obtained for $\chi'$ in the circuit simulations. The linear dispersive readout, however, has low fidelity. Similarly, when both the linear and nonlinear terms are included, the results do not differ notably from the purely nonlinear readout. 

For the simulations here to be valid, it is important to verify that we are still operating within a dispersive regime of the readout. 
In other words, the resonator photon population during readout must be below the critical photon number ($n_{\text{crit}}$), which is usually designated as the photon number at which the dressed qubit states are equal superpositions of the original qubit states~\cite{blais2004cavity}. 
For multi-level systems, this is more complicated and requires accounting for higher-order transitions~\cite{nesterov2024measurement, dumas2024measurement,kurilovich2025high}. 
To estimate the critical photon number of the circuit in Fig.~\ref{fig:4 fluxonium molecule}(a) and to take into account the influence of higher-order transitions, we calculate the overlap in the fluxonium molecule degree of freedom between dressed states $\ket{q,0}$ and $\ket{q^{\prime},n}$, where
$q = \{0 ,1\}$, $q^{\prime} = \{1,2,3,4,5,6\}$ refer to the fluxonium molecule excitations and $n$ refers to resonator photon number. 
For photon numbers below 25, we do not observe hybridization of any qubit state with any higher level above an overlap level of 0.05, and overlaps do not exceed 0.1 until photon numbers above 60.
These statements remain true whether the circuit is balanced or imbalanced.
The detailed results for this calculation can be found in Appendix~\ref{app:n_crit}. 
We conclude that imbalancing the fluxonium molecule to achieve suppression of $\chi$ does not decrease $n_{\text{crit}}$ and that $\bar{n} \leq 15$ is below the estimated critical regime. 

Our results indicate that, despite the complexity of this fluxonium molecule circuit and the deviation from a perfect two-level circuit behavior, we can still use a mixed-coupling architecture to implement our nonlinear readout scheme with minimal residual shot-noise dephasing. 

\section{\label{sec:conclusion}Conclusion~\&~Outlook}

We considered a `mixed' spin-boson coupling regime, which refers to the simultaneous presence of bilinear couplings of non-commuting quadratures of the spin and boson being paired.
We showed that in this regime, it is possible to suppress the dispersive shift while maintaining a nonzero nonlinear qubit-state-dependent shift, the Kerr shift.
In the absence of linear response, the photon shot noise dephasing rate scales $\sim \bar{n}_{\mathrm{th}}^2$ and is therefore reduced when $\bar{n}_{\mathrm{th}}\ll 1$.
Given this improvement, we proposed a nonlinear readout scheme in which we probe the Kerr shift.

Next, we demonstrated the implementation of a mixed coupling Hamiltonian in two superconducting quantum circuits: a Cooper pair transistor and a fluxonium molecule.
Our circuit implementations achieve the desired coupling terms by changing the imbalance of the capacitive and Josephson energies of the lumped elements in the qubit. 
The fluxonium molecule instance shows that the mixed coupling concept for non-simultaneously suppressing $\chi$ and $\chi'$ works beyond the two-level approximation. 
Finally, with experimentally realistic parameters and readout power within the dispersive limit, we illustrated using the fluxonium molecule that this readout scheme can be used to achieve high-fidelity readout with suppression of residual photon shot noise.
A possible experimental challenge in implementing the fluxonium molecule circuit for nonlinear readout is the flux crosstalk. Further work could explore possible circuits in which a single-mode qubit is coupled to the readout resonator with mixed coupling. 

Finally, we consider another application of our concept:  the spin serving as ancilla for the bosonic mode, which can serve as a quantum memory~\cite{cai2021bosonic,sivak2023real,liu2024hybrid}. 
One of the principal issues of such architectures is the reciprocal problem of the one described here: spin errors are inherited as uncorrectable errors in the boson.  
Because the leading-order interaction is typically the dispersive shift $\chi$, $T_1$-type spin errors are inherited as uncorrectable bosonic dephasing, which is concurrently precisely the noise channel that such cavities without qubits are meant to be biased against~\cite{reagor2016quantum, liu2024hybrid}
To combat this, various quantum control approaches have been developed to use reduced values of $\chi$~\cite{eickbusch2022fast, diringer2024conditional}. 
Our concept suggests an alternative approach, whereby linear bosonic dephasing is reduced by nullifying $\chi$ entirely while accomplishing quantum control through the Kerr shift interaction.
We leave a technical evaluation of this idea for future work.

\section*{Author contributions}
V.F.~initiated the project.
V.F.~and A.L.R.M.~developed the theory of dispersive shift suppression in a two-level system.
C.K.A.~and V.F.~proposed using mixed-coupling for nonlinear readout.
All authors contributed to the design of the circuit implementations, and A.L.R.M.,~J.H., and A.M. performed the circuit simulations.
J.H.,~T.S.,~and C.K.A. performed the readout simulations.
J.H.,~A.L.R.M.,~C.K.A.,~and V.F.~wrote the manuscript.

\begin{acknowledgments}

The authors thank Isidora Araya-Day, Anton Akhmerov, and Baptiste Royer for useful discussions.
J.H. and C.K.A acknowledge funding support from NWO Open Competition Science M, and T.V.S. acknowledges support from the Engineering and Physical Sciences Research Council (EP-SRC) under EP/SO23607/1.
A.L.R.M.~acknowledges the funding support from NWO VIDI Grant 016.Vidi.189.180.
A.L.R.M.~and V.F.~acknowledge funding from Global Quantum Leap, National Science Foundation AccelNet Program, Award No. OISE-2020174.

\end{acknowledgments}

\onecolumngrid

\appendix

\section{Perturbation theory on the mixed-coupling Hamiltonian}
\label{app:lowdin}

We perform a Schrieffer-Wolff transformation on the Hamiltonian from Eq.~(\ref{eq:generic_spin_boson}) symbolically.
The code to reproduce our calculations is available in~\cite{zenodo_code}.
The leading-order correction to the Hamiltonian:
\begin{align}
    \delta H =\frac{\hbar}{\omega_r^2 - \omega_q^2}\left[ \omega_r g_P g_X (X^2 + P^2) - \omega_q (g_P^2 P^2 + g_X^2 X^2) \right]\sigma_z~.
    \label{eq:lowest_order}
\end{align}
Note that the couplings $g_X$ and $g_P$ result in qubit-state-dependent corrections to the inductance $\sim X^2$ and capacitance $\sim P^2$ of the resonator.
From Eq.~(\ref{eq:lowest_order}), we extract the dispersive shift by collecting the terms proportional to $a^{\dagger}a \sigma_z$, resulting in the expression shown in Eq.~(\ref{eq:analytical_chi}).
We can find the roots of Eq.~(\ref{eq:analytical_chi}) analytically, giving the following relation between $g_P$ and $g_X$:
\begin{equation}
    g_P = \frac{g_X}{\omega_q}\left(\omega_r \pm \sqrt{ \omega_r^2 -\omega_q^2}\right)~.
    \label{eq:zero-condition}
\end{equation}

In Fig.~\ref{fig:tls}~(b, c), we show the resulting dispersive and Kerr shifts from a symbolic Schrieffer-Wolff transformation up to the next-leading-order correction.
Our expression for $\chi=0$ in Eq.~(\ref{eq:zero-condition}) still holds since we are deep in the perturbative regime, and therefore, the next-leading-order corrections to $\chi$ are negligible (note that $\chi^{\prime}/\chi \ll 1$).
However, the full expressions are available in~\cite{zenodo_code}.

\section{\texorpdfstring{Dephasing rate due to $\chi'$}{Dephasing rate due to chi prime}}
\label{app:dephasing}

References \cite{clerk2007using} and \cite{rigetti2012superconducting} both investigated qubit dephasing into a thermal resonator.
However, in both works, only the linear dispersive shift is considered.
Here we derive the qubit dephasing rate due to the coupling term $\chi' \sigma_z \hat{a}^\dagger \hat{a}^\dagger \hat{a} \hat{a}$ using a similar approach as used in~\cite{rigetti2012superconducting}. 

We consider the following dispersive Hamiltonian
\begin{align}
\frac{H}{\hbar} = \omega_{r} \hat{a}^\dagger \hat{a} 
+ \left( \chi'_{0} \ket{0}\bra{0} + \chi'_{1} \ket{1}\bra{1} \right) \hat{a}^\dagger \hat{a}^\dagger \hat{a} \hat{a}
+ \tilde{\omega}_q \ket{1}\bra{1},
\label{eq:Hamiltonian-qubit-resonator}
\end{align}
where $\tilde{\omega}_q$ is the Lamb shifted qubit frequency, $\chi'_{0}$ and $\chi'_{1}$ are the qubit state-dependent shifts on the resonator anharmonicity. Here, we only investigate qubit dephasing due to thermal photons, so we assume there is no active driving through the resonator. We describe the effects of having thermal photon populations in the resonator with the master equation
\begin{align}
\dot{\rho} &= -\frac{i}{\hbar} [H, \rho] 
+ \kappa (n_{\text{th}} + 1) \mathcal{D}[\hat{a}] \rho 
+  \kappa n_{\text{th}} \mathcal{D}[\hat{a}^\dagger] \rho,
\label{eq:master-eq}
\end{align}where $\mathcal{D}[\hat{\mathcal{L}}]\rho = (2\mathcal{\hat{L}}\rho\mathcal{\hat{L}}^\dagger - \mathcal{\hat{L}}^\dagger \mathcal{\hat{L}} \rho - \rho \mathcal{\hat{L}}^\dagger \mathcal{\hat{L}})/2$ is the superoperator that describes the effect of dissipation, $\kappa$ is the linewidth of the resonator, and $n_{\text{th}}$ is the thermal photon number in the resonator. 

We decompose the system to write the qubit-resonator density matrix as
\begin{align}
\rho &= \rho^{11} \ket{1}\bra{1} + \rho^{00} \ket{0}\bra{0} 
+ \rho^{10} \ket{1}\bra{0} + \rho^{01} \ket{0}\bra{1},
\label{eq:qubit-res-density-matrix}
\end{align} where $\rho^{xy}$ only acts in the resonator Hilbert space. Substituting Eqs.~(\ref{eq:Hamiltonian-qubit-resonator}) and (\ref{eq:qubit-res-density-matrix}) into the master equation, we get
\begin{align}
\dot{\rho}^{11} &= \kappa (n_{\text{th}} +1)\mathcal{D}[\hat{a}] \rho^{11} + \kappa n_{\text{th}} \mathcal{D}[\hat{a}^\dagger] \rho^{11} -i \omega_r[\hat{a}^\dagger \hat{a}, \rho^{11}] - i \chi_1' [\hat{a}^\dagger \hat{a}^\dagger \hat{a} \hat{a}, \rho^{11}],\\
\dot{\rho}^{00} &= \kappa (n_{\text{th}} + 1) \mathcal{D}[\hat{a}] \rho^{00} + \kappa n_{\text{th}} \mathcal{D}[\hat{a}^\dagger] \rho^{00} -i \omega_r[\hat{a}^\dagger \hat{a}, \rho^{00}] - i \chi_0' [\hat{a}^\dagger \hat{a}^\dagger \hat{a} \hat{a,} \rho^{00}], \\
\dot{\rho}^{10} &= \kappa (n_{\text{th}} + 1) \mathcal{D}[\hat{a}] \rho^{10} + \kappa n_{\text{th}} \mathcal{D}[\hat{a}^\dagger] \rho^{10} - i \omega_r [\hat{a}^\dagger \hat{a}, \rho^{10}] 
+ i \chi_0' \rho^{10} \hat{a}^\dagger \hat{a}^\dagger \hat{a} \hat{a} - i \chi_1' \hat{a}^\dagger \hat{a}^\dagger \hat{a} \hat{a} \rho^{10},
\label{eq.differential_rho10}
\end{align}
and $\rho^{01}=(\rho^{10})^*$.

To solve the differential equations above, we express the density matrix using the positive-P representation~\cite{2004_Gardiner_Book}:
\begin{equation}
    \rho = \int d^2\alpha \int d^2\beta \frac{|\alpha\rangle \langle \beta^*|}{\langle \beta^*|\alpha \rangle} P(\alpha, \beta)~,
    \label{eq.positive-P}
\end{equation}
and use the identities 
\begin{align}
    \hat{a} \frac{|\alpha\rangle \langle \beta^*|}{\langle \beta^*|\alpha \rangle} = \alpha \frac{|\alpha\rangle \langle \beta^*|}{\langle \beta^*|\alpha \rangle}~,\quad \quad
    \hat{a}^\dagger \frac{|\alpha\rangle \langle \beta^*|}{\langle \beta^*|\alpha \rangle} = (\beta + \partial_\alpha) \frac{|\alpha\rangle \langle \beta^*|}{\langle \beta^*|\alpha \rangle}~, \\
    \frac{|\alpha\rangle \langle \beta^*|}{\langle \beta^*|\alpha \rangle} \hat{a}^\dagger = \beta \frac{|\alpha\rangle \langle \beta^*|}{\langle \beta^*|\alpha \rangle}~, \quad \quad
    \frac{|\alpha\rangle \langle \beta^*|}{\langle \beta^*|\alpha \rangle} \hat{a} = (\alpha + \partial_\beta) \frac{|\alpha\rangle \langle \beta^*|}{\langle \beta^*|\alpha \rangle}~.
\end{align}
Applying these identities to Eq.~(\ref{eq.positive-P}), using partial integration and assuming that the boundary terms vanish ($P(\pm \infty,\pm \infty)=0$), we obtain the following correspondences. For details, see Sec. 6.4 of \textit{Quantum Noise}~\cite{2004_Gardiner_Book}.
\begin{align}
    \hat{a}\rho \leftrightarrow \alpha P(\alpha,\beta)~, \quad
    \hat{a}^\dagger \rho \leftrightarrow  (\beta - \partial_\alpha)P(\alpha,\beta)~, \quad
   \rho \hat{a}^\dagger \leftrightarrow \beta P(\alpha,\beta)~, \quad 
    \rho \hat{a} \leftrightarrow  (\alpha - \partial_\beta) P(\alpha,\beta)~.
\end{align}

Since we want to determine the qubit dephasing rate, we focus only on the $\rho^{10}$ differential equation. The positive-P representation of Eq.~(\ref{eq.differential_rho10}) gives 
\begin{align}
    \dot{P}^{10} &= \partial_\alpha [(i \omega_r +\kappa/2)\alpha P^{10}] + \partial_\beta [(-i \omega_r +\kappa/2)\beta P^{10}] \nonumber +\kappa n_{\text{th}} \partial^2_{\alpha,\beta}P^{10} \notag \\&\quad  + i \chi'_0(\alpha -  \partial_\beta)^2\beta^2 P^{10} - i \chi'_1(\beta - \partial_\alpha)^2\alpha^2 P^{10}~.
\end{align}
To solve this differential equation, we introduce the transformation
\begin{equation}
    P(\alpha, \beta) = \int da \int db \bar{P}(a,b) e^{ia\alpha+ib\beta+c.c}~,
\end{equation}
which leads to the identities:
\begin{align}
    \partial_\alpha P(\alpha, \beta) &= \int da \int db \, (i a) \overline{P}(a, b) e^{i a \alpha + i b \beta + \text{c.c.}}, \\
    \partial_\beta P(\alpha, \beta) &= \int da \int db \, (i b) \overline{P}(a, b) e^{i a \alpha + i b \beta + \text{c.c.}}, \\
    \alpha P(\alpha, \beta) &= \int da \int db \, (i) \partial_a \overline{P}(a, b) e^{i a \alpha + i b \beta + \text{c.c.}}, \\
    \beta P(\alpha, \beta) &= \int da \int db \, (i) \partial_b \overline{P}(a, b) e^{i a \alpha + i b \beta + \text{c.c.}}. 
\end{align}
The resulting differential equation for $\bar{P}^{10}(a,b)$ after the transformation is:
\begin{align}
    \partial_t \bar{P}^{10} &= \kappa/2 (-a\partial_a-b\partial_b)\bar{P}^{10} + i \omega_r (-a\partial_a+b\partial_b)\bar{P}^{10} \notag - \kappa n_{\text{th}} ab\bar{P}^{10} \\&\quad+i\chi'_0 (\partial_{a} -b)^2\partial_{b}^2\bar{P}^{10}-i\chi'_1 (\partial_{b} - a)^2\partial_a^2\bar{P}^{10}.
    \label{eq.ODE_P}
\end{align}

We consider thermal driving only and start from a Gaussian ansatz for $\bar{P}^{10}(a,b)$. Since the nonlinear terms in Eq.~(\ref{eq.ODE_P}) generate contributions proportional to $a^2 b^2$, a purely Gaussian form is not preserved by the dynamics. This motivates the inclusion of an additional factor $e^{a^2 b^2 W(t)}$, leading to the extended ansatz
\begin{equation}
\bar P^{10}(a,b,t)
=
e^{i\mu(t)}
\exp\!\left[
-\frac{Z(t)}{2}\,ab + W(t)a^2b^2
\right].
\label{eq:extended_ansatz_W}
\end{equation}
Substituting this ansatz into Eq.~(\ref{eq.ODE_P}) yields
\begin{equation}
i\dot\mu-\frac{\dot Z}{2}ab+\dot W a^2b^2
=
\left(\frac{\kappa Z}{2}-\kappa n_{\mathrm{th}}\right)ab
-2\kappa W a^2b^2
-2i\chi'\mathcal F(ab).
\label{eq:master_projected}
\end{equation}

where 

\begin{align}
\mathcal F(ab) &=\sum_{m=0}^{6} C_m (ab)^m, \\
C_0 &= 4W+\frac{Z^2}{2},\\
C_1 &= -16WZ-8W-\frac{Z^3}{2}-Z^2,\\
C_2 &= 68W^2+9WZ^2+14WZ+2W+\frac{Z^4}{16}+\frac{Z^3}{4}+\frac{Z^2}{4},\\
C_3 &= -48W^2Z-40W^2-WZ^3-3WZ^2-2WZ,\\
C_4 &= 80W^3+6W^2Z^2+12W^2Z+4W^2,\\
C_5 &= -16W^3Z-16W^3,\\
C_6 &= 16W^4.
\end{align}

\begin{figure}[t]
    \centering
    \includegraphics[width=0.9\linewidth]{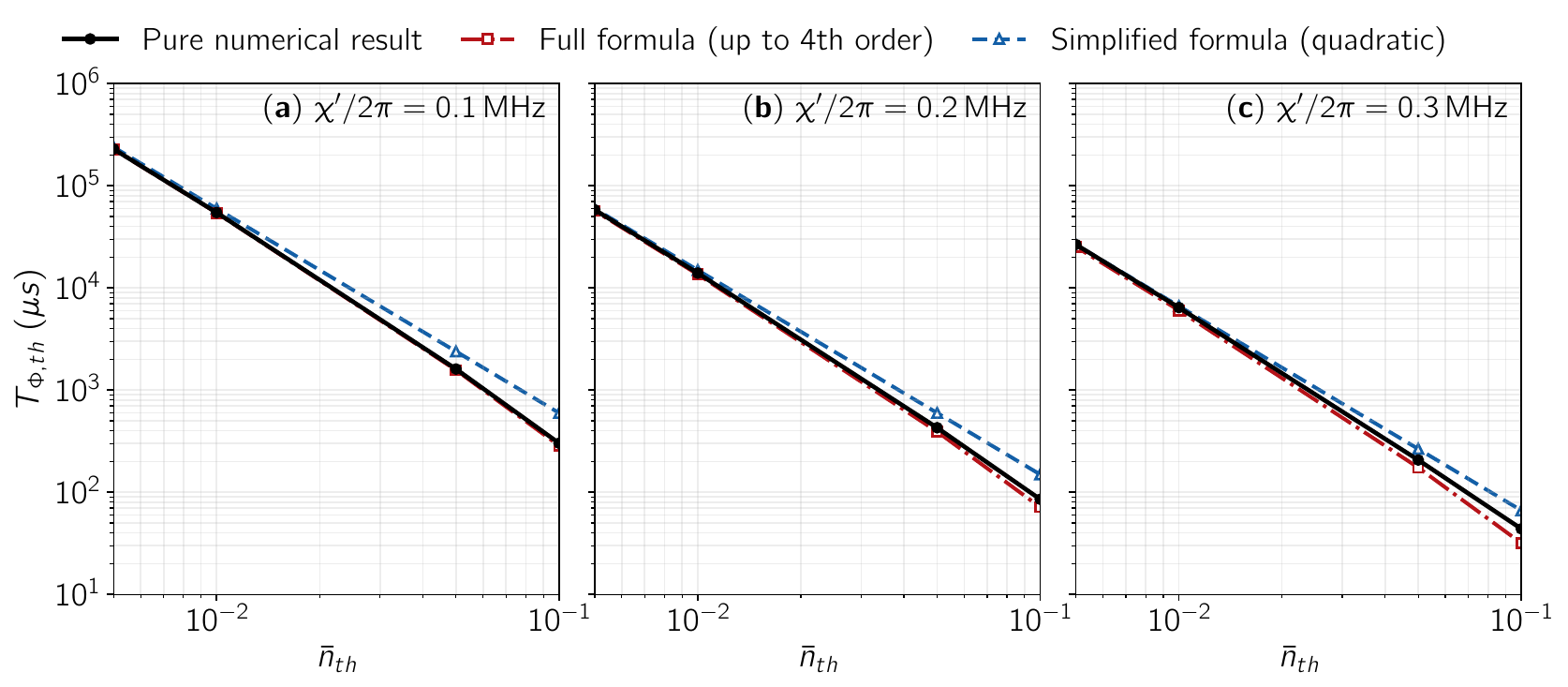}
    \caption{Nonlinear shot noise dephasing as a function of $\bar{n}_{\mathrm{th}}$ with $\kappa/2\pi = 3$ MHz for three different values of $\chi'$.The black solid lines are numerical simulations. The red dashed line is from~Eq. (\ref{eq:Gamma_final_W}). The blue dashed line uses~Eq. (\ref{eq:Gamma_low_n_W}).}
    \label{fig:dephasing_numerics}
\end{figure}

The left-hand side contains only the structures $1$, $ab$, and $(ab)^2$, whereas $\mathcal F(ab)$ contains terms up to $(ab)^6$. Hence, the ansatz Eq. \eqref{eq:extended_ansatz_W} does \emph{not} close exactly under the dynamics. We project Eq.~\eqref{eq:master_projected} onto the subspace spanned by $\{1, ab,(ab)^2\}$ and match coefficients of the retained powers. To validate this truncation, we compare the dephasing rate extracted from the reduced equations with direct numerical solutions of the full master equation and find good agreement in the regime $\chi'/\kappa\ll 1$ and $n_{\mathrm{th}}\ll 1$ \cite{zenodo_code}, see~Fig. \ref{fig:dephasing_numerics}.

Matching coefficients after the truncation gives
\begin{align}
\dot{\mu} &=
-\chi' Z^2 - 8\chi' W,
\label{eq:mu_dot_Z}
\\
\dot{Z} &=
-\kappa Z + 2\kappa n_{\mathrm{th}}
- 2i\chi'
\left(
Z^3 + 2Z^2 + 32WZ + 16W
\right),
\label{eq:Z_dot_Z}
\\
\dot{W} &=
-2\kappa W
- 2i\chi'
\left(
68W^2 + 9WZ^2 + 14WZ + 2W + \frac{Z^4}{16} + \frac{Z^3}{4} + \frac{Z^2}{4}
\right).
\label{eq:W_dot_W}
\end{align}
where $\chi^\prime=(\chi^\prime_1-\chi^\prime_0)/2$.

To extract the long-time dephasing rate, we consider the weak-nonlinearity regime
\begin{equation}
\chi' \ll \kappa.
\end{equation}
At $\chi'=0$, Eqs.~\eqref{eq:Z_dot_Z} and \eqref{eq:W_dot_W} give the stationary solution
\begin{equation}
Z^{(0)}=2n_{\mathrm{th}},
\qquad
W^{(0)}=0.
\end{equation}
We write the first-order corrections
\begin{equation}
Z=2n_{\mathrm{th}}+\delta Z,
\qquad
W=\delta W,
\end{equation}
with $\delta Z,\delta W=O(\chi'/\kappa)$.

Keeping only terms of order $O(\chi')$, one finds the stationary values
\begin{align}
Z_\ast &=
2n_{\mathrm{th}}
-\frac{16i\chi'}{\kappa}\,n_{\mathrm{th}}^2(n_{\mathrm{th}}+1)
+O(\chi'^2),
\label{eq:Z_star}
\\
W_\ast &=
-\frac{i\chi'}{\kappa}\,n_{\mathrm{th}}^2(n_{\mathrm{th}}+1)^2
+O(\chi'^2).
\label{eq:W_star}
\end{align}
Substituting Eqs.~\eqref{eq:Z_star} and \eqref{eq:W_star} into Eq.~\eqref{eq:mu_dot_Z} and 
expanding to order $O(\chi'^2)$, we obtain 
\begin{equation}
\dot\mu_\ast
=
-4\chi' n_{\mathrm{th}}^2
+\frac{8i\chi'^2}{\kappa}
n_{\mathrm{th}}^2(n_{\mathrm{th}}+1)(9n_{\mathrm{th}}+1)
+O(\chi'^3).
\label{eq:mu_dot_star_final}
\end{equation}
We parameterize the coherence factor as
\begin{equation}
e^{i\mu(t)}
=
\exp\!\left[-\int_0^t \bigl(\Gamma(t')+i\Delta(t')\bigr)\,dt'\right],
\end{equation}
where $\Gamma(t)$ and $\Delta(t)$ are the instantaneous dephasing rate and frequency shift, respectively. In the long-time limit, since $\dot{\mu}(t)\to\dot{\mu}_\ast$, we have
\begin{equation}
\Gamma=\operatorname{Im}(\dot{\mu}_\ast),
\qquad
\Delta=-\operatorname{Re}(\dot{\mu}_\ast).
\end{equation}
Therefore, the dephasing rate is
\begin{equation}
\Gamma
=
\frac{8\chi'^2}{\kappa}\,
n_{\mathrm{th}}^2(n_{\mathrm{th}}+1)(9n_{\mathrm{th}}+1)
+O(\chi'^3).
\label{eq:Gamma_final_W}
\end{equation}
In the low-temperature limit $n_{\mathrm{th}}\ll 1$, this reduces to
\begin{equation}
\Gamma \simeq \frac{8\chi'^2}{\kappa}\,n_{\mathrm{th}}^2.
\label{eq:Gamma_low_n_W}
\end{equation}

\section{Readout simulations and validations}
\label{app:ro_sim}

Optimal linear weights are applied to the extraction of the SNR from the semi-classical equation of motion (Eq. (\ref{langevin_resonator})). The readout error is expressed as  
\begin{equation}
P_{\mathrm{err}}(t_m)
=
\frac{1}{2}\,\operatorname{erfc}\!\left(
\frac{1}{2}\sqrt{
\eta\kappa
\int_0^{t_m} |\alpha_0(t')-\alpha_1(t')|^2\,dt'
}
\right).
\label{eq:perr-compact}
\end{equation}

For comparison with the semiclassical treatment, we numerically solve the full Lindblad master equation for the density matrix $\rho$ of the qubit-resonator system using \texttt{Qiskit Dynamics}~\cite{Puzzuoli2023},
\begin{equation}
\frac{d\rho}{dt}
=
-i\,[H(t),\rho]
+
\sum_{j=1}^{2}
\left(
L_j \rho L_j^\dagger
-
\frac{1}{2}\left\{L_j^\dagger L_j,\rho\right\}
\right).
\end{equation}
We decompose the Hamiltonian as
\begin{equation}
H(t)=H_{\mathrm{sys}}+H_{\mathrm{drive}}(t),
\end{equation}
with
\begin{align}
H_{\mathrm{sys}}
&=
\omega_{\mathrm{r}}\left(\mathbb{I}_q \otimes a^\dagger a\right)
+
\frac{\omega_q}{2}\left(\sigma_z \otimes \mathbb{I}_r\right)
+
\chi'\left(\sigma_z \otimes a^{\dagger 2} a^2\right),
\\[4pt]
H_{\mathrm{drive}}(t)
&=
s(t)\, i\left(\mathbb{I}_q \otimes a^\dagger - \mathbb{I}_q \otimes a\right).
\end{align}
The dissipative dynamics are described by the jump operators
\begin{align}
L_1
&=
\sqrt{\kappa(n_{\mathrm{th}}+1)}\left(\mathbb{I}_q \otimes a\right),
\\[4pt]
L_2
&=
\sqrt{\kappa n_{\mathrm{th}}}\left(\mathbb{I}_q \otimes a^\dagger\right).
\end{align}

The resonator Hilbert space is truncated to $N_r=50$ Fock states. To maintain consistency with the drive convention used in Eq.~\eqref{langevin_resonator}, the master-equation simulation is implemented with a square pulse
\begin{equation}
s(t)=
\begin{cases}
2\epsilon, & 0 \le t \le t_{\mathrm{ro}},\\
0, & \text{otherwise}.
\end{cases}
\end{equation}
The evolution is implemented in the resonator rotating frame, with carrier frequency matched to $\omega_{\mathrm{r}}/2\pi$. The prefactor of $2$ in the square pulse amplitude appears because the master-equation solver uses a real lab-frame drive, whereas Eq.~\eqref{langevin_resonator} is written in terms of the co-rotating drive amplitude. In the rotating frame of the resonator, the resonant part of the lab-frame drive $2\epsilon \cos(\omega_{\mathrm{r}} t)$ yields the effective amplitude $\epsilon$, matching the semiclassical equation. 

To compare directly with the semiclassical result, we extract the qubit-conditioned resonator amplitudes from the master-equation evolution as expectation values of the resonator annihilation operator, and use these trajectories in the same SNR and readout-error expression as in the semiclassical case, Eq.~\eqref{eq:perr-compact}.

\subsection{Linewidth dependence}

\begin{figure}[t]
    \centering
    \includegraphics[width=\linewidth]{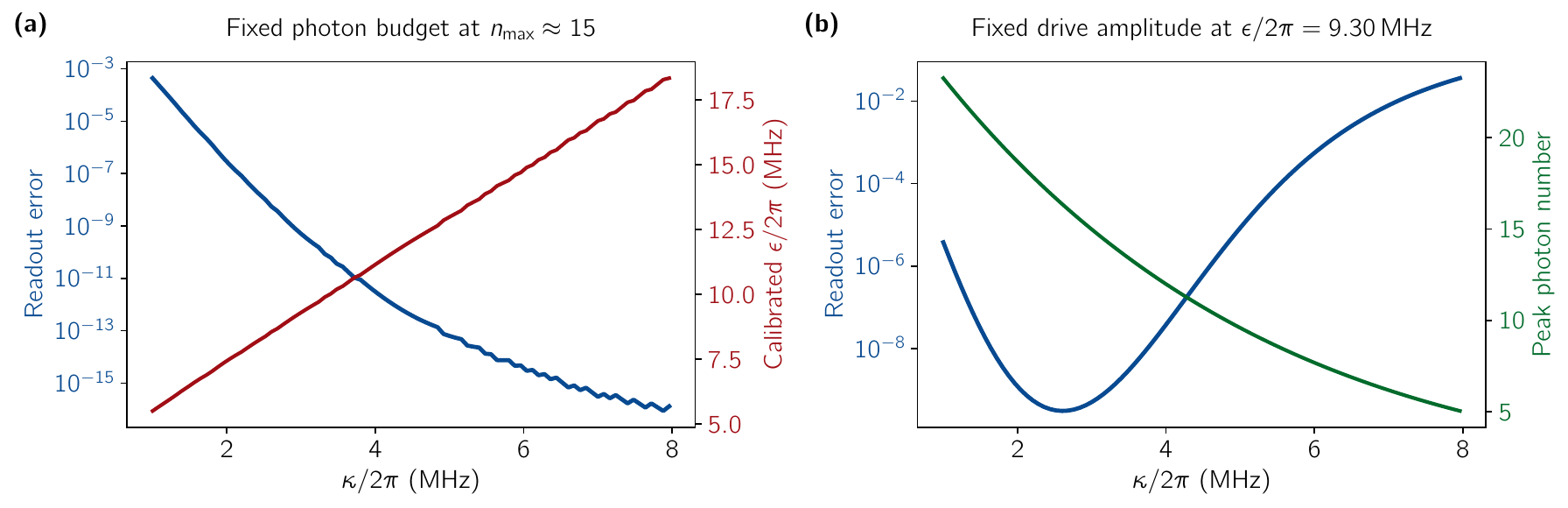}
    \caption{Nonlinear readout error as a function of readout linewidth \(\kappa\) with \(\chi'/2\pi=0.12~\mathrm{MHz}\). In both panels, the integration time is \(200~\mathrm{ns}\). In (a), we calibrate the drive amplitude such that the peak intracavity photon number remains approximately fixed, whereas (b) is obtained at a fixed drive amplitude.}
    \label{fig:errorlinewidth}
\end{figure}

In the nonlinear readout scheme, varying the linewidth \(\kappa\) affects both the cavity response time and, through the intracavity photon number, the effective strength of the nonlinear pull \(\chi_{\mathrm{eff}} \sim 2\chi'|\alpha|^2\).

Fig.~\ref{fig:errorlinewidth} compares two linewidth sweeps for the case \(\chi'/2\pi=0.12~\mathrm{MHz}\), with integration time of \(t_m=200~\mathrm{ns}\). In panel (a), the drive amplitude is recalibrated for each \(\kappa\) such that the peak intracavity photon number remains approximately fixed. Over the scanned range, the error decreases monotonically with increasing \(\kappa\), reflecting the faster cavity response and more rapid extraction of measurement information. In panel (b), by contrast, the drive amplitude is held fixed. In this case, varying \(\kappa\) also changes the photon population and hence the effective nonlinear pull. The readout error therefore becomes nonmonotonic, with a minimum near \(\kappa/2\pi \approx 3\text{--}4~\mathrm{MHz}\), reflecting a competition between faster information leakage and weaker nonlinear contrast at larger \(\kappa\).

In the simulations used throughout this work, we choose \(\kappa/2\pi=3~\mathrm{MHz}\) as a representative operating point. This value gives low readout error while avoiding unnecessarily large Purcell-limited qubit relaxation.

\subsection{Dispersive regime validation}

\label{app:n_crit}
\begin{figure}
    \centering
    \includegraphics[width=0.65\linewidth]{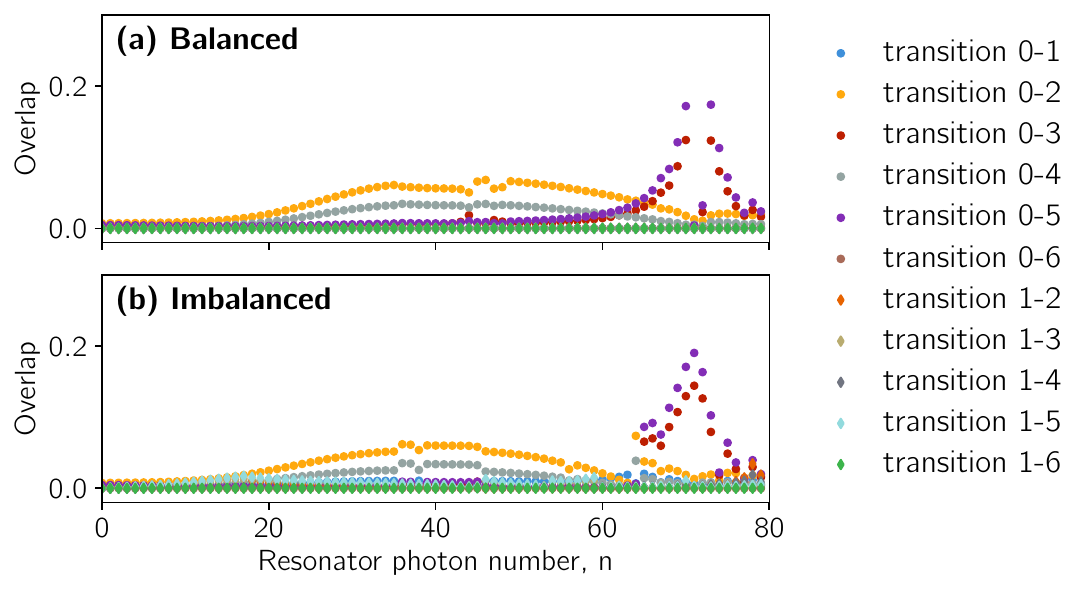}
    \caption{Qubit state overlap as a function of resonator photon number. The balanced one (a) serves as a reference and has the same circuit parameters as in Fig.~\ref{fig:4 fluxonium molecule}(c) except that the fluxonium molecule is perfectly symmetric (i.e, both the capacitance and the Josephson energy imbalance are zero). The imbalanced one (b) has the same circuit parameters used for the readout simulation in Fig.~\ref{fig:4 fluxonium molecule}(d).}
    \label{fig:overlap_n}
\end{figure}

To validate the dispersive treatment underlying the readout model, we use exact diagonalization of the full circuit Hamiltonian obtained from \texttt{scqubits}. We label the dressed eigenstates as $\ket{q,n}$ and calculate the overlap of the qubit degree of freedom between states $\ket{q,0}$ and $\ket{q',n}$ for possible qubit transition $i-j$. More concretely, we calculate the fidelity between the reduced qubit states $\text{Tr}_{res} \rho_{\ket{q,0}}$ and $\text{Tr}_{res} \rho_{\ket{q',n}}$ as a function of resonator photon number $n$. Fig.~\ref{fig:overlap_n} (a) shows the case with an unsuppressed $\chi/2\pi$ of $2$~MHz and $\chi'/2\pi=0.05$~MHz, where we see no abrupt hybridization until $n\approx 70$. A similar trend is found for the imbalanced case where $\chi$ is suppressed. Fig.~\ref{fig:overlap_n} (b) features $\chi/2\pi=-0.02$~MHz and $\chi'/2\pi=-0.12$~MHz.

\section{Mixed coupling with a Cooper-pair transistor}
\label{app:cpt-derivation}
In this appendix, we derive the mixed qubit--resonator
coupling generated by the Cooper-pair transistor circuit shown
in Fig.~\ref{fig:CPT_circuit_and_results}. We distinguish physical node fluxes
$\Phi_j$, which have units of magnetic flux, from the
corresponding dimensionless superconducting phases $\phi_j$:
\begin{equation}
    \phi_j = \frac{\Phi_j}{\varphi_0},
    \qquad
    \varphi_0 \equiv \frac{\Phi_0}{2\pi}
    = \frac{\hbar}{2e}.
    \label{eq:cpt-reduced-flux-quantum}
\end{equation}
The circuit Lagrangian is $\mathcal{L}=T-U$, with
\begin{align}
T &=
    \frac{C}{2}
    \left(\dot{\Phi}_2-\dot{\Phi}_1\right)^2
    +
    \frac{C_{J1}}{2}
    \left(\dot{\Phi}_J-\dot{\Phi}_1\right)^2+
    \frac{C_{J2}}{2}
    \left(\dot{\Phi}_J-\dot{\Phi}_2\right)^2,
\\
U &=
    \frac{
        \left(
            \Phi_2-\Phi_1-\Phi_{\mathrm{ext}}
        \right)^2
    }{2L}
    -
    E_{J1}\cos\left(\phi_J-\phi_1\right)
    -
    E_{J2}\cos\left(\phi_J-\phi_2\right).
\end{align}
We assume that the external
flux is static.

We introduce the resonator coordinate $\phi_R = \phi_2-\phi_1-\phi_{\mathrm{ext}}$ and the island coordinate $\phi_I = \phi_J-(\phi_1+\phi_2)/2$. The kinetic energy becomes

\begin{align}
T =
\frac{\varphi_0^2}{2}
\Bigg[
    C\dot{\phi}_R^{\,2}
    &+
    C_{J1}
    \left(
        \dot{\phi}_I+\frac{\dot{\phi}_R}{2}
    \right)^2+
    C_{J2}
    \left(
        \dot{\phi}_I-\frac{\dot{\phi}_R}{2}
    \right)^2
\Bigg].
\label{eq:cpt-kinetic-new}
\end{align}
Defining
\begin{equation}
    E_{J\Sigma}=E_{J1}+E_{J2},
    \qquad
    E_{J\Delta}=E_{J1}-E_{J2},
    \label{eq:cpt-EJ-definitions}
\end{equation}
the potential energy can be written as
\begin{align}
U ={}&
    \frac{\varphi_0^2}{2L}\phi_R^2
    -
    E_{J\Sigma}
    \cos\left(
        \frac{\phi_{\mathrm{ext}}+\phi_R}{2}
    \right)
    \cos\phi_I +  E_{J\Delta}
    \sin\left(
        \frac{\phi_{\mathrm{ext}}+\phi_R}{2}
    \right)
    \sin\phi_I.
\end{align}

We write the kinetic energy $T = \mathcal{L}_C = \frac{\varphi_0^2}{2}\dot{\phi}^T\mathbf{C}\dot{\phi}$, so the capacitance matrix in the coordinate basis
$\boldsymbol{\phi}=(\phi_R,\phi_I)^T$ is
\begin{equation}
    \mathbf{C}
    =
    \begin{pmatrix}
        C+\dfrac{C_{J1}+C_{J2}}{4}
        &
        \dfrac{C_{J1}-C_{J2}}{2}
        \\[2mm]
        \dfrac{C_{J1}-C_{J2}}{2}
        &
        C_{J1}+C_{J2}
    \end{pmatrix},
    \label{eq:cpt-capacitance-matrix}
\end{equation}
where we see in the off-diagonal elements the origin of the capacitive coupling between the Cooper pair transistor mode and the resonator mode. 
For later convenience, we define
\begin{align}
    C_{J\Sigma}
    &=
    C_{J1}+C_{J2},
    &
    C_{J\Delta}
    &=
    C_{J1}-C_{J2},
\end{align}

The canonical momenta conjugate to the dimensionless phases satisfy
\begin{equation}
    p_j
    =
    \frac{\partial\mathcal{L}}{\partial\dot{\phi}_j}
    =
    \hbar n_j,
    \qquad
    [\phi_j,n_k]
    =
    i\delta_{jk},
\end{equation}
While the corresponding physical charges are
\begin{equation}
    Q_j=2en_j.
\end{equation}
Including the island offset charge $n_g$, the charging Hamiltonian is
\begin{equation}
    H_C
    =
    \frac{1}{2}
    \boldsymbol{Q}^{\,T}
    \mathbf{C}^{-1}
    \boldsymbol{Q},
    \qquad
    \boldsymbol{Q}
    =
    2e
    \begin{pmatrix}
        n_R\\
        n_I-n_g
    \end{pmatrix}.
\end{equation}
We therefore obtain
\begin{align}
H_C ={}&
    4E_{CR}n_R^2
    +
    4E_{CI}(n_I-n_g)^2
    -
    4E_{C\times}n_R(n_I-n_g).
    \label{app:charging_H}
\end{align}
where
\begin{align}
    E_{CR}=
    \frac{e^2}{2}
    \frac{C_{J\Sigma}}{\det\mathbf{C}}, \quad 
    E_{CI}=
    \frac{e^2}{2}
    \frac{
        C+C_{J\Sigma}/4
    }{\det\mathbf{C}}, \quad 
    E_{C\times}
    =
    \frac{e^2}{2}
    \frac{C_{J\Delta}}{\det\mathbf{C}}.
    \label{eq:cpt-ECdelta-effective}
\end{align}
Notice that $E_{C\times}$ here is an effective cross-charging energy obtained from the inverse capacitance matrix. In general, it is not equal to the difference between the individual junction charging
energies.
\subsection{Cooper Pair Box limit}
\label{app:cpt-cpb-limit}

We now consider the Cooper-pair-box regime, in which the charging
energy is sufficiently large that the island dynamics can be
described using a small number of charge states. Near the charge
degeneracy point, we retain the two adjacent charge states
$\{\lvert 0\rangle_n,\lvert 1\rangle_n\}$. The Josephson coupling to
all other charge states is assumed to be perturbatively small
compared with their charging-energy separation.

Within this two-dimensional charge subspace, we write the
charge-basis Pauli operators as $\tau_i$, where $i\in \{x,y,z\}$ and make substitutions
\begin{equation}
    n_I-n_g
    \longrightarrow
    -\frac{1}{2}
    \left(
        \tau_z+n_g'\mathbb I
    \right), \qquad
        \cos\phi_I
    \longrightarrow
    \frac{\tau_x}{2},
    \qquad
    \sin\phi_I
    \longrightarrow
    -\frac{\tau_y}{2}.
    \label{eq:cpt-projection}
\end{equation}

where we defined the detuning from the charge-degeneracy point as
\begin{equation}
    n_g'
    \equiv
    2n_g-1,
    \label{eq:cpt-ng-prime}
\end{equation}
Starting from the charging Hamiltonian in
Eq.~\eqref{app:charging_H}, the two-level projection gives
\begin{align}
H_C ={}&
    4E_{CR}n_R^2
    +
    2E_{CI}n_g'\tau_z
    +
    2E_{C\times}n_R
    \left(
        \tau_z+n_g'\mathbb I
    \right)
    +
    \mathrm{const.}
    \label{eq:cpt-charging-two-level}
\end{align}
Here, the additive constant
$E_{CI}(1+n_g'^2)$ has been omitted.

The Josephson potential becomes
\begin{align}
U_J ={}&
    -
    \frac{E_{J\Sigma}}{2}
    \cos\left(
        \frac{\phi_{\mathrm{ext}}+\phi_R}{2}
    \right)
    \tau_x
    -
    \frac{E_{J\Delta}}{2}
    \sin\left(
        \frac{\phi_{\mathrm{ext}}+\phi_R}{2}
    \right)
    \tau_y.
    \label{eq:cpt-Josephson-two-level}
\end{align}

Assuming that the resonator phase fluctuations are small, we expand
the Josephson terms to second order in $\phi_R$. We define
\begin{equation}
    c
    \equiv
    \cos\left(\frac{\phi_{\mathrm{ext}}}{2}\right),
    \qquad
    s
    \equiv
    \sin\left(\frac{\phi_{\mathrm{ext}}}{2}\right).
    \label{eq:cpt-cs-definitions}
\end{equation}
The Josephson contribution then takes the form
\begin{align}
U_J ={}&
    -
    \frac{E_{J\Sigma}}{2}
    c\,\tau_x
    -
    \frac{E_{J\Delta}}{2}
    s\,\tau_y
    +
    \frac{\phi_R}{4}
    \left(
        E_{J\Sigma}s\,\tau_x
        -
        E_{J\Delta}c\,\tau_y
    \right)
    +
    \frac{\phi_R^2}{16}
    \left(
        E_{J\Sigma}c\,\tau_x
        +
        E_{J\Delta}s\,\tau_y
    \right)
    +
    \mathcal O(\phi_R^3).
    \label{eq:cpt-Josephson-expanded}
\end{align}

Combining the charging, inductive, and Josephson contributions, the
full Hamiltonian can be written as
\begin{equation}
    H
    =
    H_R+H_I+H_{\mathrm{coupling}},
    \label{eq:cpt-H-decomposition}
\end{equation}
with
\begin{align}
    H_R 
    &=
    4E_{CR}n_R^2
    +
    2E_{C\times}n_g'n_R
    +
    \frac{\varphi_0^2}{2L}\phi_R^2, \\
    H_I
    &=
    2E_{CI}n_g'\tau_z
    -
    \frac{E_{J\Sigma}}{2}c\,\tau_x
    -
    \frac{E_{J\Delta}}{2}s\,\tau_y,
    \\
    H_{\mathrm{coupling}} &=
    2E_{C\times}n_R\tau_z+
    \frac{\phi_R}{4}
    \left(
        E_{J\Sigma}s\,\tau_x
        -
        E_{J\Delta}c\,\tau_y
    \right)+
    \frac{\phi_R^2}{16}
    \left(
        E_{J\Sigma}c\,\tau_x
        +
        E_{J\Delta}s\,\tau_y
    \right).
    \label{eq:Hcharge-basis}
\end{align}

We introduce the dimensionless parameters
\begin{align}
    r
    &\equiv
    \frac{E_{J\Delta}}{E_{J\Sigma}},
    \quad
    z\equiv
    -c+irs, 
    \quad 
    m\equiv
    \frac{4E_{CI}n_g'}{E_{J\Sigma}}, 
    \quad
    \varepsilon
    \equiv
    \sqrt{m^2+|z|^2}.
    \label{eq:cpt-dimensionless-definitions}
\end{align}
The static island Hamiltonian can then be written as
\begin{equation}
    H_I
    =
    \frac{E_{J\Sigma}}{2}
    \left(
        -c\,\tau_x
        -
        rs\,\tau_y
        +
        m\,\tau_z
    \right).
    \label{eq:cpt-Hq-vector}
\end{equation}
Equivalently, in the charge basis,
\begin{equation}
    H_I
    =
    \frac{E_{J\Sigma}}{2}
    \begin{pmatrix}
        m & z\\
        z^* & -m
    \end{pmatrix}.
    \label{eq:cpt-Hq-matrix}
\end{equation}
Its eigenenergies are $E_{\pm}=\pm E_{J\Sigma}\varepsilon/2$.

The charge states $\lvert 0\rangle_n$ and $\lvert 1\rangle_n$ are
generally not eigenstates of $H_I$ because they are mixed by the
Josephson terms. For $\lvert z\rvert\neq0$, we introduce the unitary
transformation
\begin{equation}
    R
    =
    \sqrt{\frac{\lvert z\rvert^2}{2\varepsilon}}
    \begin{pmatrix}
        \dfrac{\sqrt{\varepsilon+m}}{\lvert z\rvert}
        &
        -\dfrac{\sqrt{\varepsilon-m}}{\lvert z\rvert}
        \\[3mm]
        \dfrac{\sqrt{\varepsilon-m}}{z}
        &
        \dfrac{\sqrt{\varepsilon+m}}{z}
    \end{pmatrix}.
    \label{eq:cpt-basis-rotation}
\end{equation}
whose columns are the excited and ground states, respectively,
expressed in the charge basis. Denoting the Pauli operators in the
energy eigenbasis by $\sigma_i$, we have
\begin{equation}
    H_I^{(E)} = R^\dagger H_I R
    =
    \frac{E_{J\Sigma}\varepsilon}{2}\sigma_z,
    \qquad
    O^{(E)}=R^\dagger O R.
    \label{eq:cpt-Hq-energy-basis}
\end{equation}
Applying the same transformation to
Eq.~\eqref{eq:Hcharge-basis} gives
\begin{align}
H_{\mathrm{coupling}}^{(E)}
={}&
\Bigg[
    \frac{2E_{C\times}m}{\varepsilon}n_R
    -
    \frac{
        E_{J\Sigma}^2-E_{J\Delta}^2
    }{
        8E_{J\Sigma}\varepsilon
    }
    \sin(\phi_{\mathrm{ext}})\phi_R
    -
    \frac{E_{J\Sigma}|z|^2}{16\varepsilon}\phi_R^2
\Bigg]\sigma_z
\nonumber\\
&+
\Bigg[
    -
    \frac{2E_{C\times}|z|}{\varepsilon}n_R
    -
    \frac{
        m\left(E_{J\Sigma}^2-E_{J\Delta}^2\right)
    }{
        8E_{J\Sigma}\varepsilon|z|
    }
    \sin(\phi_{\mathrm{ext}})\phi_R
    -
    \frac{E_{J\Sigma}m|z|}{16\varepsilon}\phi_R^2
\Bigg]\sigma_x
\nonumber\\
&+
\frac{E_{J\Delta}}{4|z|}\phi_R\sigma_y.
\label{eq:cpt-interaction-energy-basis}
\end{align}
At the charge-degeneracy point, $n_g'=0$, so that
$m=0$ and $\varepsilon=\lvert z\rvert$. Equation
\eqref{eq:cpt-interaction-energy-basis} then reduces to
\begin{align}
H_{\mathrm{coupling}}^{(E)}
={}&
\Bigg[
    -
    \frac{
        E_{J\Sigma}^2-E_{J\Delta}^2
    }{
        8E_{J\Sigma}|z|
    }
    \sin(\phi_{\mathrm{ext}})\phi_R
    -
    \frac{E_{J\Sigma}|z|}{16}\phi_R^2
\Bigg]\sigma_z
\nonumber\\
&-
2E_{C\times}n_R\sigma_x
+
\frac{E_{J\Delta}}{4|z|}\phi_R\sigma_y.
\label{eq:cpt-charge-degeneracy-coupling}
\end{align}
Thus, at charge degeneracy, the only off-diagonal terms to this order
are the two desired mixed-quadrature couplings; the remaining terms
are diagonal in the qubit energy basis.

\section{Detailed circuit simulation results}
\label{app:circ-simulation-details}

\begin{figure}
    \centering    \includegraphics[width=\linewidth]{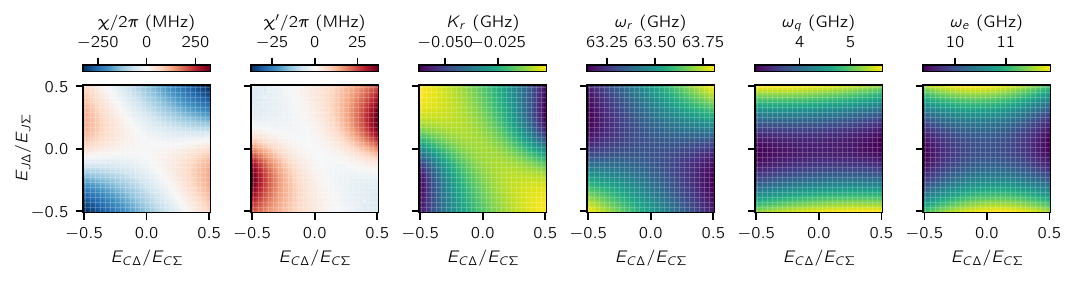}
    \caption{A more detailed simulation result showing the change of $\chi$, $\chi'$, $K_r$, $\omega_r$, $\omega_q$, and the transition frequency of the Cooper pair transistor from the ground state to the second excited state $\omega_e$ as a function of the capacitance and Josephson energy imbalances.
    }
    \label{fig:chi-kerr-cpt}
\end{figure}

\begin{figure}
    \centering    \includegraphics[width=\linewidth]{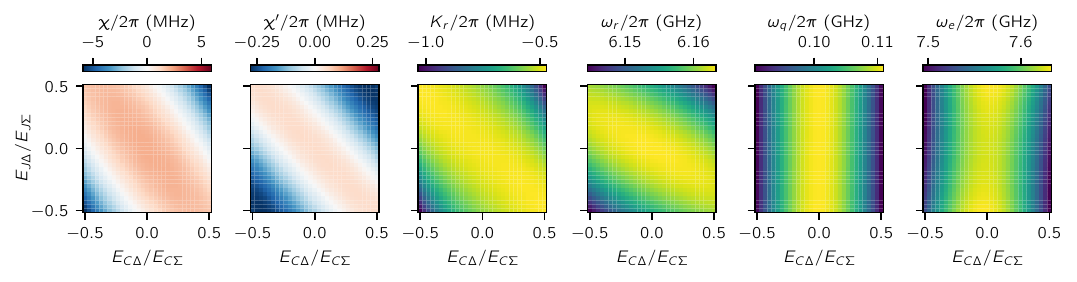}
    \caption{A more detailed simulation result showing the change of $\chi$, $\chi'$, $K_r$, $\omega_r$, $\omega_q$, and the transition frequency of the fluxonium molecule from the ground state to the second excited state $\omega_e$ as a function of the capacitance and Josephson energy imbalances.
    }
    \label{fig:chi-kerr-fx}
\end{figure}

Here we provide additional numerical simulations of the circuits shown in Fig.~\ref{fig:CPT_circuit_and_results}(a) and Fig.~\ref{fig:4 fluxonium molecule}(a).
We repeat the dispersive and Kerr shift plots from the main text and additionally show the Kerr shift of the resonator $K_r$ and the frequencies of the resonator ($\omega_r$), qubit ($\omega_q$), and first excited state in the qubit ($\omega_e$).
The fact that there are no sudden jumps in these energies demonstrates that the qubit dispersive and Kerr shifts are suppressed due to the mixed coupling rather than the qubit and the resonator level crossings.
Additionally, we show that the Kerr shift of the resonator remains finite, see Fig.~\ref{fig:chi-kerr-cpt} and Fig.~\ref{fig:chi-kerr-fx}.

\begin{figure}
    \centering
    \includegraphics[width=0.65\linewidth]{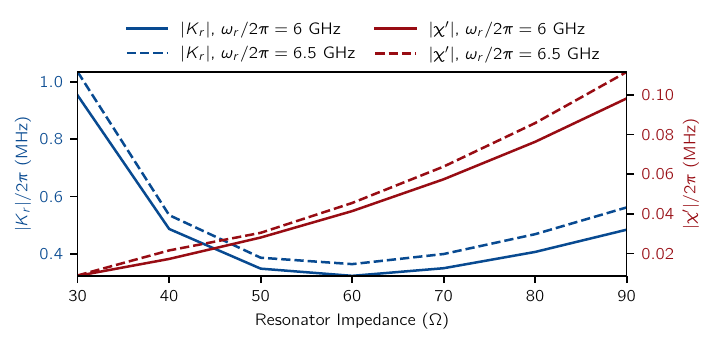}
    \caption{Resonator self-Kerr $K_r$ and Kerr shift $\chi'$ as a function of resonator impedance for fixed resonator frequencies of $6$ GHz and $6.5$~GHz. The fluxonium molecule parameters are the same as in Fig.~\ref{fig:4 fluxonium molecule}(c). The Josephson energies are imbalanced such that $\chi$ is suppressed. All $\chi'$ values presented here correspond to when $|\chi|/2\pi$ is suppressed to be below $0.2$~MHz.}
    \label{fig:selfKerr}
\end{figure}

As we search for a set of realistic circuit parameters for the fluxonium molecule circuit, we find that for a wide choice of qubit parameters, independent suppression of $\chi$ and $\chi'$ can be found. However, some resonator impedance engineering is needed to suppress $K_r$ while keeping $\chi'$ high enough. In Fig.~\ref{fig:selfKerr} we see that for the circuit in Fig.~\ref{fig:4 fluxonium molecule}(a), $|K_r|$ has a local minimum at around $50~\Omega$, whereas $|\chi'|$ increases with resonator impedance. This motivates our choice of the relatively high resonator impedance in the main text.

\bibliography{biblio_with_urls}

\end{document}